\documentclass[reprint, amsmath, amssymb, superscriptaddress]{revtex4-2}

\usepackage{graphicx}
\graphicspath{{./plots/}}
\usepackage{natbib}
\usepackage{amsmath}
\usepackage[font={small, it}]{caption}
\usepackage{float}
\usepackage{dcolumn}
\usepackage{bm}
\usepackage[colorlinks=true,citecolor=blue,linkcolor=blue,urlcolor=blue]
{hyperref}

\begin{document}
	
	\title{The role of closed timelike curves in particle motion within Van Stockum space-time: A generalization}
	
	
	\author{Ayanendu Dutta}
	\email{ayanendudutta@gmail.com}
	\affiliation{Department of Physics, Jadavpur University, Kolkata-700032, INDIA}
		
	\author{Dhritimalya Roy}
	\email{rdhritimalya@gmail.com}
	\affiliation{Department of Physics, Jadavpur University, Kolkata-700032, INDIA}
	
	\author{Subenoy Chakraborty}
	\email{schakraborty.math@gmail.com}
	\affiliation{Department of Mathematics, Jadavpur University, Kolkata-700032, INDIA}

	
\begin{abstract}
	The present work analyses the particle motion in the Van Stockum space-time considering the existence of closed timelike curves. Test particles with or without angular momentum are studied in the present geometry. It is found that only non-zero angular momentum test particles exist in the neighborhood of closed timelike curves. The minimum particle energy and the range of backward time jump in closed timelike geodesics are studied within a Cauchy horizon. Finally, a general prescription for CTC and CTG has been presented with appropriate examples.
\end{abstract}
\maketitle

\section{Introduction}\label{introduction}
Causality violation in space-times is an interesting outcome of Einstein’s General Relativity. It can lead to the intersection of worldline in a previous point, and so closed timelike curves (CTC) occur.

The idea of backward time travel came into existence when G\"odel claimed that his cosmological solution \cite{RevModPhys.21.447} contains causality violating region. But Chandrasekhar and Wright proved that there is no such closed timelike geodesic (CTG) orbit present \cite{Chandrasekhar1961}. Later, Stein described that a non-geodesic closed timelike curve exists in the solution \cite{doi:10.1086/288328}. It is established that without the presence of CTG, CTC can occur.

Earlier Van Stockum described the exact solution of Field Equation for an axially symmetric infinitely long rotating dust cylinder \cite{van.stockum.1938}. The space-time is divided into an interior solution of the dust ($r<R$), and a vacuum exterior solution ($r>R$). Here '$R$' is the radius of dust cylinder, and '$r$' is the radial distance from the axis. Tipler found the possibility of a causality violating region in the solution \cite{PhysRevD.9.2203}. It was found that $ar>1$ of the interior solution, and $aR>1/2$ of the exterior solution contain closed timelike orbits ('$a$' stands for the angular velocity of the dust) \cite{Lorentzian.Wormholes.Visser} \cite{lobo2010closed}. It was not so clear whether the curve is geodesic or not. The first solution containing CTG was presented by Soares \cite{Soares1980}. And later, Steadman showed that the exterior solution of Van Stockum with $aR>1/2$ contains CTG with the circular timelike geodesic \cite{Steadman2003}.

Bonnor presented the solution of a rotating dust cloud that admits CTC \cite{Bonnor_1977}. And the absence of CTG in the metric was studied by Collas and Klein \cite{Collas2004}.

Although the study of closed timelike orbit was going on, the limit of backward time jump or the energy of the particles orbiting the curve was unknown. Matt Visser presented the backward time jump for a null azimuthal curve which in general neither necessarily be a closed null curve nor be a geodesic \cite{Lorentzian.Wormholes.Visser}. The formula is utterly important to the closed curve study, but it does not shed any light on the backward time jump in CTC or CTG.

Opher et al. studied the confinement of geodesic motion in Van Stockum interior solution \cite{doi:10.1063/1.531489}. In a specific case, the geodesic confinement is established in the radial direction where the axial movement is taken to be free. It was also observed that the behaviour of particle motion is affected by the space-time dragging.
Having these in mind the study of particle motion around closed timelike orbits is put forward in the present work. 

The question begins with, “Does every kind of particle traverse in a closed timelike path? Or should they have some kind of special characteristics to do so?” That is why the nature of the particles in the $ar>1$ region of the Van Stockum interior solution is needed to be studied so that the movement of a test particle in a closed timelike path is established.

And\'reka et al. noticed the counter-rotation of closed timelike curves in the Kerr-Newman black hole \cite{andreka2008}. It turns out that CTC may not be 'caused' by frame-dragging \cite{duan2022}. So the effect of frame-dragging on particles of the neighborhood of CTC is also analyzed.

We initiated the study by reviewing the particle motion and space-time of Van Stockum dust in sections \ref{geodesic_motion} and \ref{region}. Section \ref{closed_orbit} is dedicated to the analysis of CTC and CTG in Van Stockum space-time. Results of sections \ref{geodesic_motion} and \ref{region} are extensively used here. It is found that angular momentum-less particles are forbidden in closed timelike orbits. These results and ideas are used in section \ref{gen_pres} to have a generalized picture of the nature of particle motion in CTC and CTG.

\section{Geodesic Motion in Van Stockum: A Review}\label{geodesic_motion}
The axially symmetric interior solution of the Van Stockum rotating dust cylinder is defined as,
\begin{equation}\label{eqn1}
	\begin{split}
	ds^2=-dt^2+2ar^2d \phi dt+ r^2(1-a^2r^2)d \phi ^2 \\ +e^{-a^2r^2}(dr^2+dz^2),
	\end{split}
\end{equation}
where ‘$a$' is the angular velocity of the dust, and the four coordinates $t, \phi , r, z$ range between $- \infty$ to $+ \infty$, 0 to 2$ \pi $, 0 to $+ \infty$ and $- \infty$ to $+ \infty$ respectively.
The geodesics of the space-time can readily be found with Euler-Lagrange equation. The first order geodesics are,
\begin{equation}\label{eqn2}
	\dot{t}=(1-a^2r^2)E+aP_\phi,
\end{equation}
\begin{equation}\label{eqn3}
	\dot{\phi}=\frac{P_\phi}{r^2}-aE,
\end{equation}
\begin{equation}\label{eqn4}
	\dot{z}=P_z e^{a^2r^2},
\end{equation}
\begin{equation}\label{eqn5}
	\dot{r}^2=e^{a^2 r^2} [-\epsilon+(1-a^2 r^2) E^2+2aEP_\phi -\frac{P_\phi^2}{r^2}-P_z e^{a^2 r^2}].
\end{equation}

Here $E, P_\phi, P_z$ are integration constants termed as the total energy of the test particle (will be always taken non-negative), angular momentum, and momentum along $z$ direction respectively. $\epsilon=1,0,-1$ defines timelike, null, and spacelike signatures.

Opher et al. presented the study of geodesic motion in Van Stockum's interior solution and explained the confinements \cite{doi:10.1063/1.531489}. The study of geodesics is utterly important here to describe the nature of space-time as a whole. So we will begin the discussion by presenting a review of the geodesic motion and confinements both for null and timelike trajectories. The study fits with Opher et al., but several new results are also found.

\subsection{Radial Null Geodesics}\label{radial_null}

The radial null geodesic equation is defined by, 
\begin{equation}\label{eqn6}
	\dot{r}^2=e^{a^2 r^2} [(1-a^2 r^2 ) E^2+2aEP_\phi-\frac{P_\phi^2}{r^2}-P_z e^{a^2 r^2}].
\end{equation}

To obtain the space-time diagram of $r-t$, we apply equation (\ref{eqn2}) and (\ref{eqn5}) in the following form,
\begin{equation}\label{eqn7}
	\begin{split}
		& \frac{dt}{dr}=\frac{dt/d\lambda}{dr/d\lambda}=\frac{\dot{t}}{\dot{r}} \\& = \pm \frac{e^{-\frac{a^2 r^2}{2}} [(1-a^2 r^2 )E+aP_\phi]}{[-\epsilon + (1-a^2 r^2) E^2+2aEP_\phi -\frac{P_\phi^2}{r^2} -P_z e^{a^2 r^2}]^\frac{1}{2}}
	\end{split}
\end{equation}

We present solutions with two different kinds of combinations in angular momentum $P_\phi=0$ and $P_\phi\neq0$. Another assumption is that momentum along the $z$ direction is zero ($P_z=0$) i.e. the coaxial movements of the particles are neglected.

\subsubsection{$P_\phi=0$}\label{null_zero-ang}
For $P_\phi=0$, $P_z=0$ and $\epsilon=0$ equation (\ref{eqn7}) becomes,
\begin{equation}\label{eqn8}
	\frac{\dot{t}}{\dot{r}}=\pm e^{- \frac{a^2 r^2}{2}} \sqrt{1-a^2 r^2}.
\end{equation}

\begin{figure}[h]
	
	\centerline{\includegraphics[scale=.7]{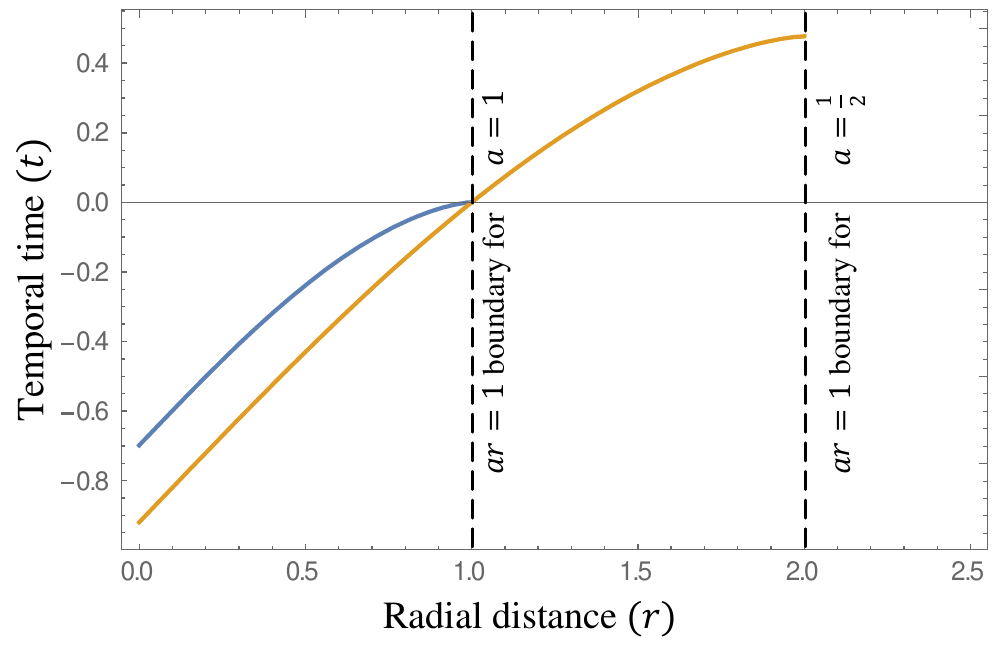}}
	\caption{$r-t$ plot with $a=1$ (blue line) and $a =\frac{1}{2}$ (yellow line). Also showing the $ar=1$ boundary}
	\label{plot1}
\end{figure}

It is interesting to note that the space-time trajectory ($r-t$ diagram) of angular momentum-less photons does not depend on the energy of the particles. It will only depend on the angular velocity of the dust cylinder. Plot has been employed in figure \ref{plot1} for the solution with different values of angular velocity $a=1$ and $a= \frac{1}{2}$.

So for $P_\phi=0$ and $P_z=0$ the maximum radius of photon trajectory is always confined within $ar=1$ boundary. The acceleration of the photons which is always directed inward is zero at the axis and increases exponentially as they move outward. As they reach the $ar=1$ boundary their velocity is zero and they are drawn backward. They move like this endlessly \cite{doi:10.1063/1.531489}.

\subsubsection{$P_\phi \neq 0$}\label{null_ang}
From equation (\ref{eqn7}),
\begin{equation}\label{eqn9}
	\frac{\dot{t}}{\dot{r}}=\pm \frac{e^{- \frac{a^2 r^2}{2}} [(1-a^2 r^2 )E+aP_\phi]}{[(1-a^2 r^2) E^2+2aEP_\phi -\frac{P_\phi^2}{r^2}]^\frac{1}{2}},
\end{equation}
\begin{figure}[h!]
	
	\centerline{\includegraphics[scale=.7]{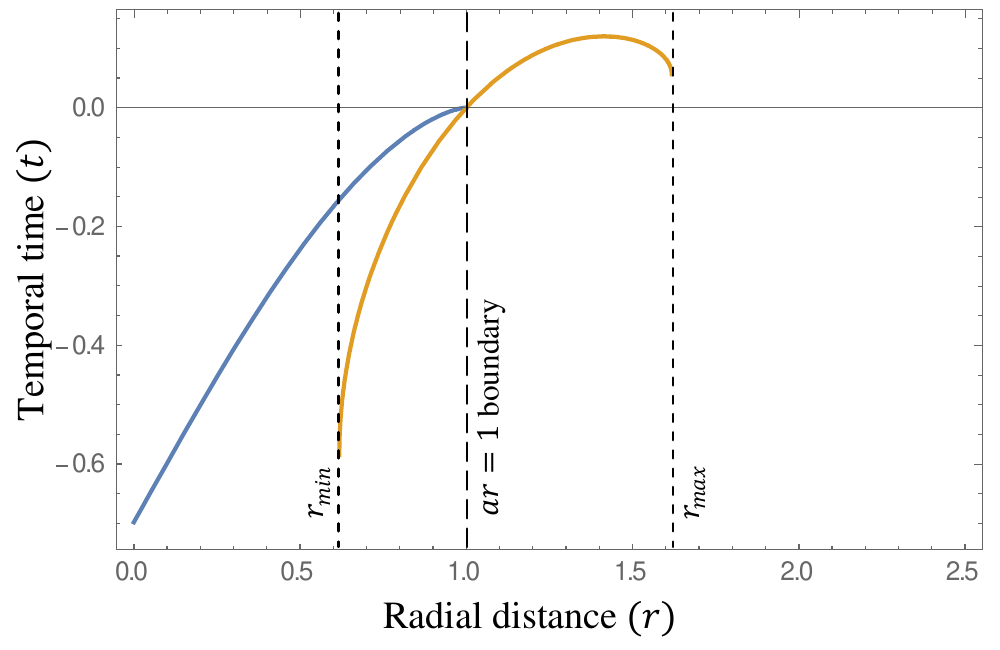}}
	\caption{$r-t$ combination plot of photon trajectory with $E=a=1, P_\phi=0$ (blue line) and $E=a=P_\phi=1$ (yellow line). The radial position of $ar=1$ boundary, $r_{min}$ and $r_{max}$ are 1, 0.618 and 1.618 respectively}
	\label{plot2}
	
	\centerline{\includegraphics[scale=.7]{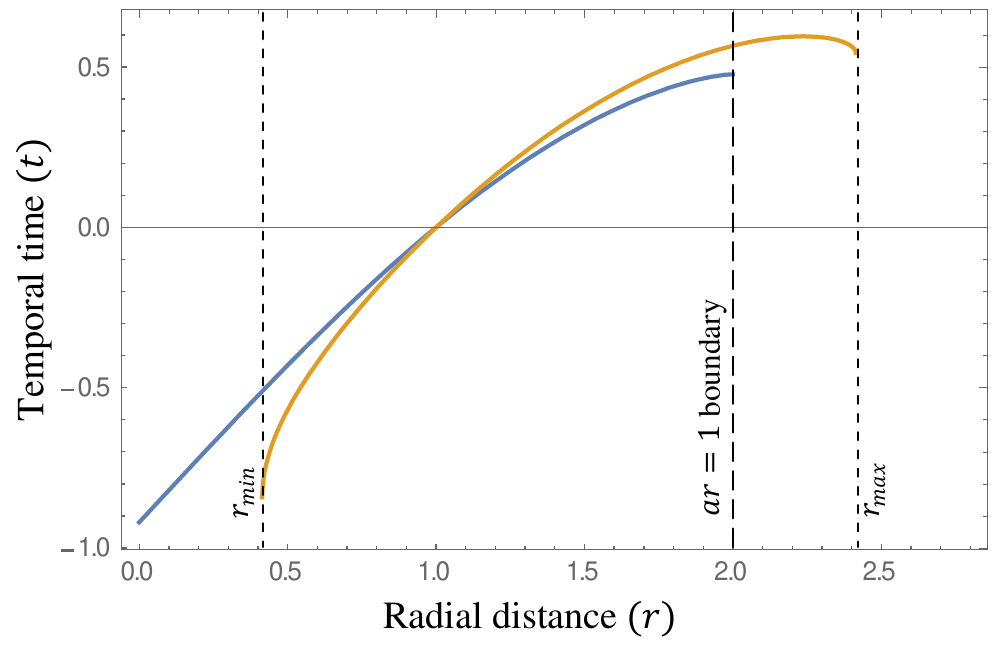}}
	\caption{$r-t$ combination plot of photon trajectory with $E=2, a=\frac{1}{2}, P_\phi=0$ (blue line) and $E=2, a=\frac{1}{2}, P_\phi=1$ (yellow line). The radial position of $ar=1$ boundary, $r_{min}$ and $r_{max}$ are 2, 0.414 and 2.414 respectively}
	\label{plot3}
\end{figure}
as the denominator should always be greater than zero, so,
\begin{equation}\label{eqn10}
	(1-a^2 r^2) E^2+2aEP_\phi -\frac{P_\phi^2}{r^2}=0,
\end{equation}
will provide the limit of the photon trajectory. It predicts two points that indicate the maximum and minimum radial reach of the photon movement. We define the points as $r_{max}$  and $r_{min}$. Therefore equation (\ref{eqn10}) gives,
\begin{equation}\label{eqn11}
	(arE- \frac{P_\phi}{r})= \pm E.
\end{equation}
Taking positive and negative value of right hand side, it respectively provides,
\begin{equation}\label{eqn12}
	r_{max}= \frac{E+\sqrt{E^2+4aEP_\phi}}{2aE},
\end{equation}
and
\begin{equation}\label{eqn13}
	r_{min}= \frac{-E+\sqrt{E^2+4aEP_\phi}}{2aE}.
\end{equation}

\begin{figure}[h!]
	
	\centerline{\includegraphics[scale=.7]{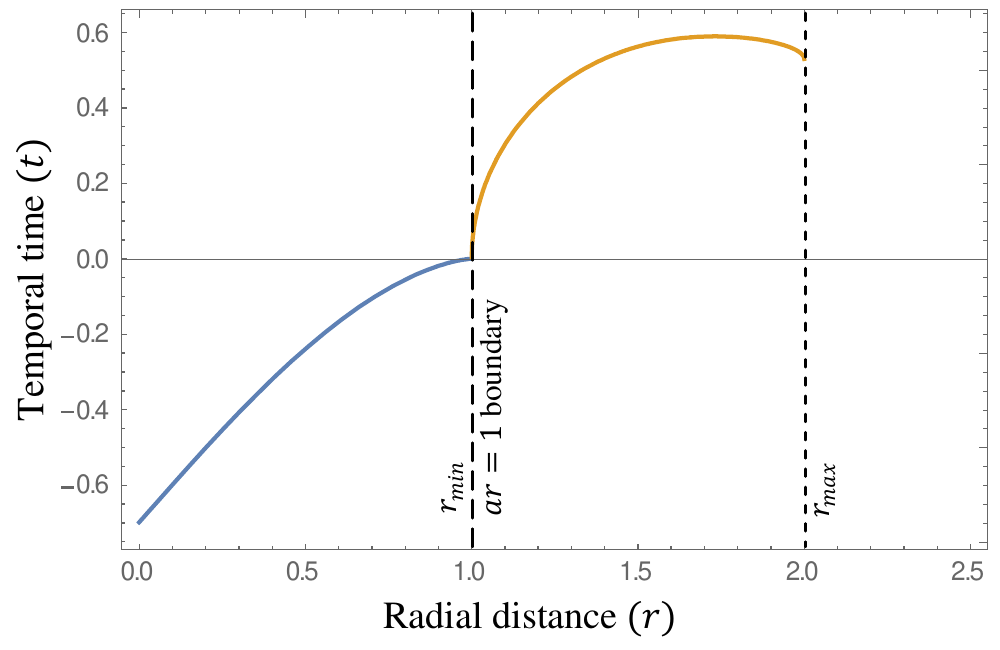}}
	\caption{$r-t$ combination plot of photon trajectory with $E=\frac{1}{2}, a=1, P_\phi=0$ (blue line) and $E=\frac{1}{2}, a=1, P_\phi=1$ (yellow line). The radial position of $ar=1$ boundary, $r_{min}$ and $r_{max}$ are 1, 1 and 2 respectively}
	\label{plot4}
	
	\centerline{\includegraphics[scale=.7]{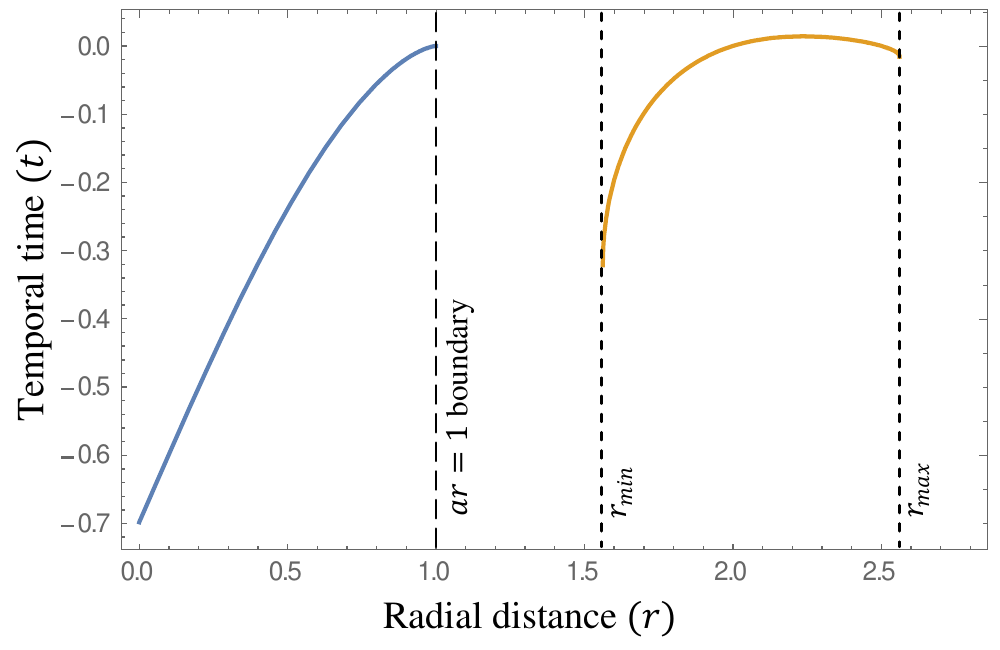}}
	\caption{$r-t$ combination plot of photon trajectory with $E=\frac{1}{4}, a=1, P_\phi=0$ (blue line) and $E=\frac{1}{4}, a=1, P_\phi=1$ (yellow line). The radial position of $ar=1$ boundary, $r_{min}$ and $r_{max}$ are 1, 1.562 and 2.562 respectively}
	\label{plot5}
\end{figure}

Figure \ref{plot2} shows a combination trajectory of $r-t$ plot of figure \ref{plot1} ($E=a=1, P_\phi=0$ in blue line) with plot of the solution of equation (\ref{eqn9}) with $E=a=P_\phi=1$ (yellow line). The $ar=1$ boundary,  $r_{max}$  and $r_{min}$ are also shown.

Figure \ref{plot2}, \ref{plot3}, \ref{plot4} and \ref{plot5} shows the combination plot of equation (\ref{eqn8}) and (\ref{eqn9}). They all show the radial position of $ar=1$ boundary, $r_{max}$ and $r_{min}$.

It is clearly noticeable that when $P_\phi \neq 0$ photon movement is bounded inside the cylindrical shell of radius $r_{min}$ and $r_{max}$ . It can also be concluded from equations (\ref{eqn6}) and (\ref{eqn9}) that at $r_{min}$ and $r_{max}$ the velocity of the photons is zero. The acceleration at the axis is infinitely large in this case and is directed outward. So the particles never reach the axis. Thus at $r_{min}$ they are pulled out before they reach further towards the axis and at $r_{max}$ they are drawn back towards the axis \cite{doi:10.1063/1.531489}.

\subsection{Radial timelike Geodesics}\label{radial_timelike}
Timelike geodesics are obtained by choosing $\epsilon=1$ in the equation (\ref{eqn5}). i.e,
\begin{equation}\label{eqn14}
	\dot{r}^2=e^{a^2 r^2} [-1+(1-a^2 r^2 ) E^2+2aEP_\phi - \frac{P_\phi^2}{r^2}-P_z e^{a^2 r^2}].
\end{equation}

Here, we discuss the massive (timelike) particle movements with two different cases $P_\phi=0$ and $P_\phi \neq 0$. We also considered the coaxial movement of the particles to be free i.e. $P_z=0$.

\subsubsection{$P_\phi=0$}\label{time_zero-ang}

For timelike particles equation (\ref{eqn7}) gives,
\begin{equation}\label{eqn15}
	\frac{\dot{t}}{\dot{r}}=\pm \frac{E e^{- \frac{a^2 r^2}{2}}(1-a^2 r^2 )}{[-1 + (1-a^2 r^2) E^2]^\frac{1}{2}}.
\end{equation}

Now for this equation to be real-valued,
\begin{equation}\label{eqn16}
	-1 + (1-a^2 r^2) E^2>0,
\end{equation}
i.e.
\begin{equation}\label{eqn17}
	E>\pm \frac{1}{\sqrt{1-a^2 r^2}}.
\end{equation}

\begin{figure}[h!]
	
	\centerline{\includegraphics[scale=.7]{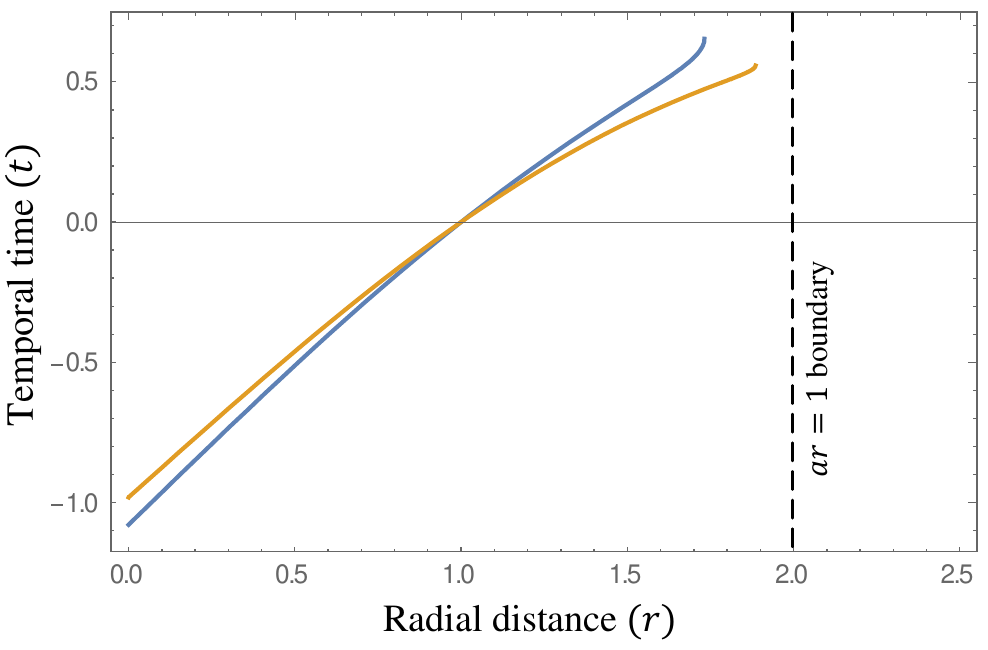}}
	\caption{$r-t$ plot of timelike test particle trajectory with $E=2, a=\frac{1}{2}$ (blue line) and $E=3, a =\frac{1}{2}$ (yellow line). Also showing the $ar=1$ boundary}
	\label{plot6}
\end{figure}

To verify this we produced two set of solutions with (i) $E=2,a=\frac{1}{2}$ and (ii) $E=3,a=\frac{1}{2}$ which are shown in $r-t$ plot of figure \ref{plot6}. It is evident from equation (\ref{eqn17}) that massive (timelike) particles are always confined within $ar<1$. Equation (\ref{eqn17}) also predicts the radial limit to be $r\approx 1.732$ and $r\approx 1.886$ respectively for (i) and (ii).

\subsubsection{$P_\phi \neq 0$}\label{time_ang}

Similar to the non-zero angular momentum photons, massive (timelike) non-zero angular momentum particles are also confined within a cylindrical shell. We define the inner and outer radius of the shell as $r_{min-time}$ and $r_{max-time}$. Here in this case equation (\ref{eqn7}) reduces to,
\begin{equation}\label{eqn18}
	\frac{\dot{t}}{\dot{r}}=\pm \frac{e^{- \frac{a^2 r^2}{2}}[(1-a^2 r^2 )E+aP_\phi]}{[-1 + (1-a^2 r^2) E^2+2aEP_\phi -\frac{P_\phi^2}{r^2}]^\frac{1}{2}}.
\end{equation}

\begin{figure}[h!]
	\centerline{\includegraphics[scale=.7]{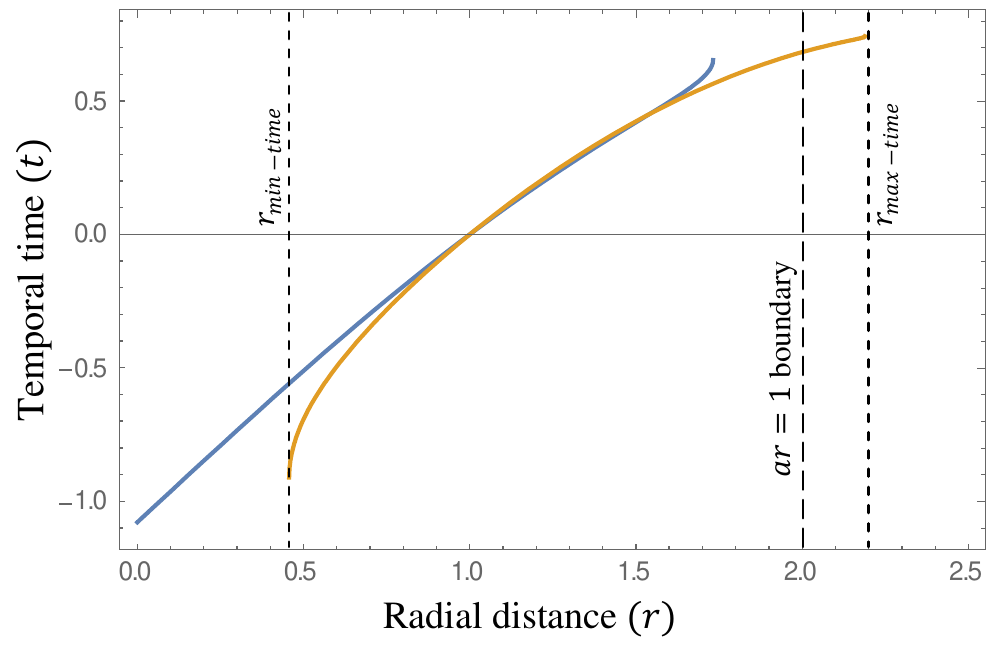}}
	\caption{Timelike test particle trajectory with $E=2, a=\frac{1}{2}$ and $P_\phi=0$ (blue line) and $E=2, a=\frac{1}{2}$ and $P_\phi=1$ (yellow line). The radial position of $ar=1$ boundary, $r_{min-time}$ and $r_{max-time}$ are 2, 0.457 and 2.189 respectively}
	\label{plot7}

	\centerline{\includegraphics[scale=.7]{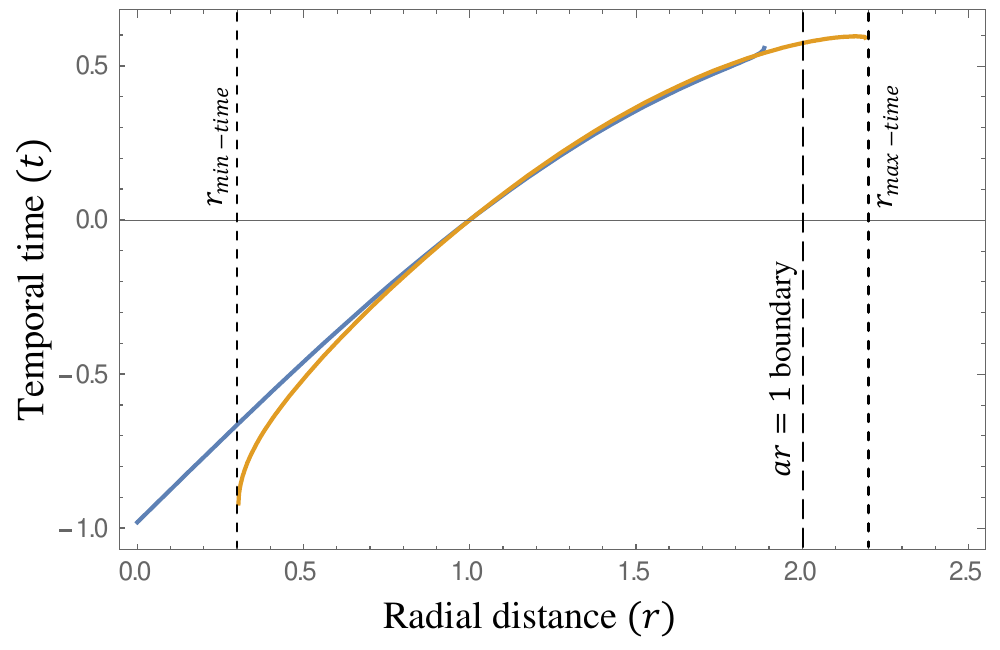}}
	\caption{Timelike test particle trajectory with $E=3, a=\frac{1}{2}$ and $P_\phi=0$ (blue line) and $E=3, a=\frac{1}{2}$ and $P_\phi=1$ (yellow line). The radial position of $ar=1$ boundary, $r_{min-time}$ and $r_{max-time}$ are 2, 0.304 and 2.19 respectively}
	\label{plot8}
\end{figure}

Now the solution to be real, the following condition must be satisfied,
\begin{equation}\label{eqn19}
	(1-a^2 r^2) E^2+2aEP_\phi -(1+ \frac{P_\phi^2}{r^2})>0,
\end{equation}
and to calculate the $r_{min-time}$ and $r_{max-time}$ we use,
\begin{equation}\label{eqn20}
	(1-a^2 r^2) E^2+2aEP_\phi -(1+ \frac{P_\phi^2}{r^2})=0,
\end{equation}
i.e.
\begin{equation}\label{eqn21}
	(arE- \frac{P_\phi}{r})=\pm \sqrt{E^2-1}.
\end{equation}

Separately taking the positive and negative values of the right hand side, the root provides $r_{max-time}$ and $r_{min-time}$ respectively as,
\begin{equation}\label{eqn22}
	r_{max-time}=\frac{\sqrt{E^2-1}+ \sqrt{E^2-1+4aEP_\phi}}{2aE},
\end{equation}
and
\begin{equation}\label{eqn23}
	r_{min-time}=\frac{- \sqrt{E^2-1}+ \sqrt{E^2-1+4aEP_\phi}}{2aE}.
\end{equation}

Two combination plots of (i) $E=2,a=\frac{1}{2}$ and (ii) $E=3,a=\frac{1}{2}$ with $P_\phi=0$ and $P_\phi=1$ are shown in figure \ref{plot7} and \ref{plot8} respectively. Equation (\ref{eqn22}) and (\ref{eqn23}) predicted the maximum and minimum radial reach of the plots as $r_{min-time}\approx 0.457$, $r_{max-time}\approx 2.189$ and $r_{min-time}\approx 0.304$, $r_{max-time}\approx 2.19$ respectively which perfectly fits with the plots. Likewise the null geodesic trajectories, the plots indicate that the non-zero angular momentum timelike test particles are also confined within cylindrical shell of radius $r_{min-time}$ to $r_{max-time}$. The $ar=1$ boundary lies in between them. As $r_{min}$ and $r_{max}$ do not coincide with $r_{min-time}$ and $r_{max-time}$, the width of the three regions for null and timelike geodesics may or may not be the same.

\section{The Regions and Boundaries}\label{region}
The existence of $ar=1$ boundary, $r_{max}$ and $r_{min}$ define multiple regions of particle trajectory. The boundaries are solely dependent and vary with the angular momentum ($P_\phi$) and the energy ($E$) of the particles. From figures \ref{plot2} and \ref{plot3} it is obvious that the interior of the dust cylinder contains three different regions, i.e.

(i) Region I: Between the axis and $r_{min}$, only angular momentum-less particles can be found.

(ii) Region II: Region between $r_{min}$ and $ar=1$ boundary can be termed as the mixed region where both zero and non-zero angular momentum particles will be present.

(iii) Region III: $ar=1$ boundary to $r_{max}$ is purely a non-zero angular momentum particle region.

The observation is also true for massive (timelike) test particles. Figures \ref{plot7} and \ref{plot8} show the same behaviour. Photons create the boundary and limit of massive particle movement. For the same reason the $r_{min-time}$ to $r_{max-time}$ region always lies inside $r_{min}$ to $r_{max}$ region.

The basic assumption of the Van Stockum interior is $r<R$, where $R$ is the radius of the dust cylinder. And outside the dust, the solution is defined by the exterior solution. So if $r_{max}$ is greater than the radius of the dust cylinder ($R$) then the interior will be joined to the exterior solution (and the whole study will be different).

Considering pure interior particle movement, we conclude that the region outside $r_{max}$ i.e. $r>r_{max}$ is a pure empty space. Even the movement of photons (and all massive particles) are completely forbidden here.
So observers sitting outside the dust cylinder (i.e. at $r>r_{max}$) can't see the interior particle movements of the cylinder. The cylinder will appear as a perfect dark cylindrical region to them.

The plots in figure \ref{plot2}, \ref{plot3}, \ref{plot4}, \ref{plot5} clearly shows that the position of $r_{min}$ and $r_{max}$ varies with different values of $E,P_\phi$ and $a$. Figure \ref{plot2} and \ref{plot3} indicate that $ar=1$ boundary always lies within $r_{min}$ to $r_{max}$ cylindrical shell. The shell gradually shifts away from the axis of the cylinder when the energy of the particle ($E$) decreases (with fixed $P_\phi$  and $a$). Regions I, II, and III all are present in these cases. But $E=\frac{1}{2}$ with $P_\phi=a=1$ is a special case (figure \ref{plot4}) where the $r_{min}$ to $r_{max}$ shell is so much shifted that $r_{min}$ coincides $ar=1$ boundary and the mixed region (Region II) vanishes. And if energy decreases further, Region I and Region III completely separate from each other. Here Region II is a forbidden region and the space in between Region I and III is a pure empty region where the existence of particles is forbidden.

\begin{figure}[h!]
	\centerline{\includegraphics[scale=.7]{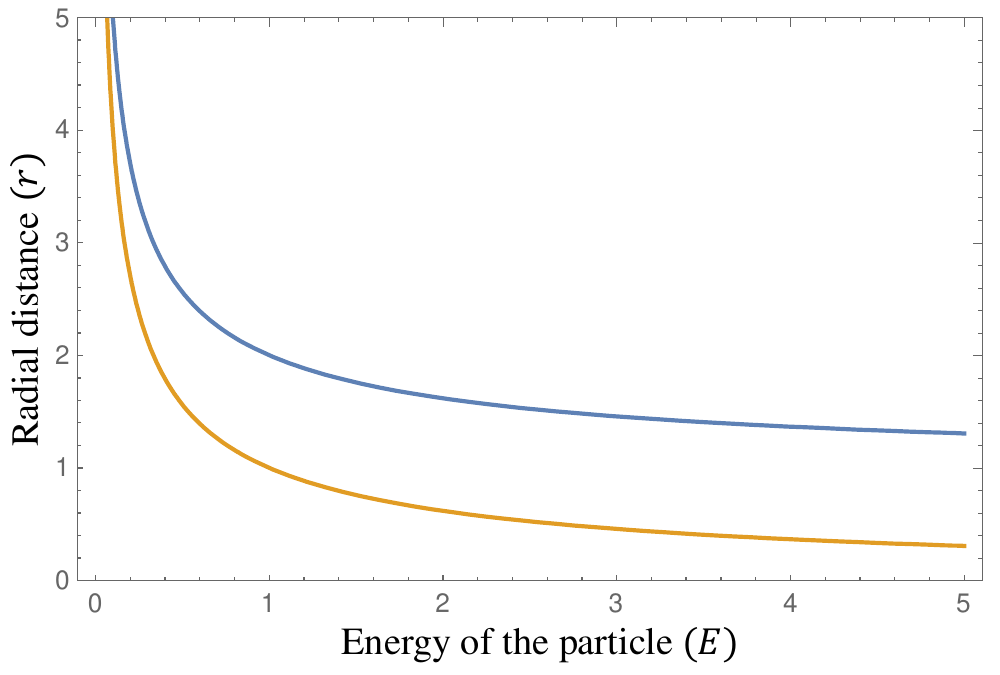}}
	\caption{Variation of $r_{min}$ with $E$ (yellow line) and $r_{max}$ with $E$ (blue line) for null geodesics (photons), where, $P_\phi=2$ and $a=1$}
	\label{plot9}
	
	\centerline{\includegraphics[scale=.7]{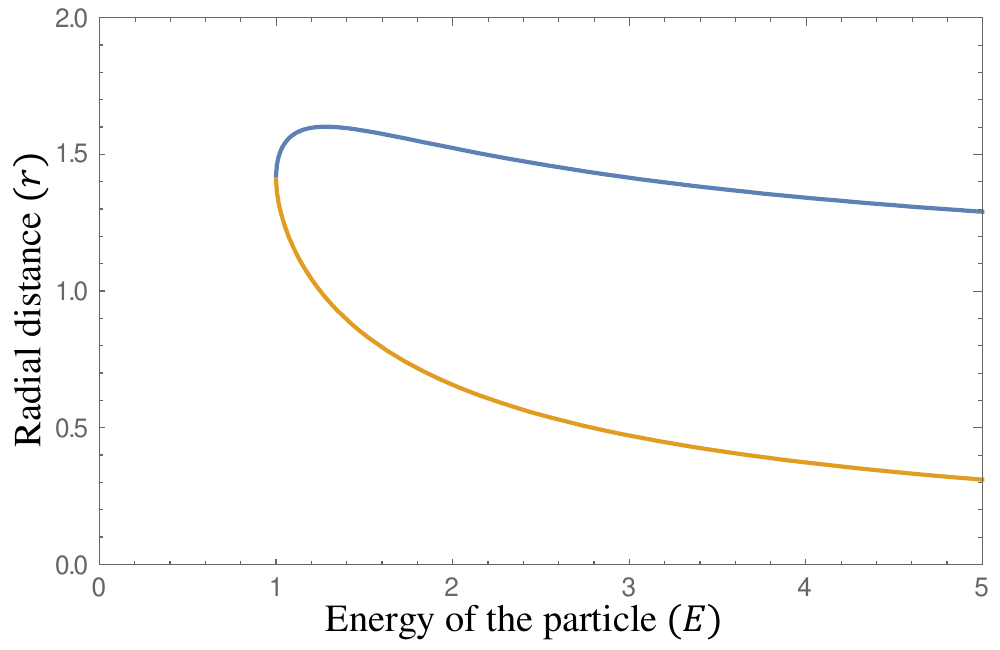}}
	\caption{Variation of $r_{min}$ with $E$ (yellow line) and $r_{max}$ with $E$ (blue line) for timelike geodesics, where, $P_\phi=2$ and $a=1$}
	\label{plot10}
\end{figure}

Now to find the relationship between $E$ and $P_\phi$ for which $r_{min}$ coincides the $ar=1$ boundary, we use equation (\ref{eqn13}). '$a$' can be taken as $\frac{1}{r}$ so that $ar=1$ always maintains. Therefore,
\begin{equation}\label{eqn24}
	r_{min}=r= \frac{-E+\sqrt{E^2+4aEP_\phi}}{2aE},
\end{equation}
i.e.
\begin{equation}\label{eqn25}
	-E+\sqrt{E^2+4EP_\phi \frac{1}{r}}=2E.
\end{equation}
So the condition arises as,
\begin{equation}\label{eqn26}
	E=\frac{P_\phi}{2r},
\end{equation}
and when $E< \frac{P_\phi}{2r}$ the Region I and Region III completely separates. Figure \ref{plot9} shows the variation of $r_{min}$ and $r_{max}$ with $E$ for $P_\phi=2$ and $a=1$.

Now, equation (\ref{eqn19}) shows the condition of a massive (timelike) particle for its movement to be physically realizable. The condition involves the values of $E, P_\phi$ and '$a$'. It is observed that massive (timelike) particle movement is forbidden for $E\le 1$ with the imposed values of $P_\phi$ and $a$. But with the same values, photons have no such limitation. So the extraordinary feature happens in the specified region where photons are present but it forbids massive particles. We can call it a 'pure null region'.

For photons, the coincidence of $r_{min}$ and $ar=1$ boundary and the separation of Region I and Region III happen at that particular condition. 

In figure \ref{plot10}, we provide the variation of $r_{min-time}$ and $r_{max-time}$ with $E$ for $P_\phi=2$ and $a=1$. It shows that at $E\le 1$ massive particle movement is forbidden. But from figure \ref{plot9} the possibility of photon movement is established. 

\subsection{Space-time dragging}\label{dragging}
Both for null and timelike test particles when they have zero angular momentum, they still have a non-zero angular velocity since equation (\ref{eqn3}) gives,
\begin{equation}\label{eqn27}
	\dot{\phi}=-aE.
\end{equation}

Opher et al. described this to be a result of space-time dragging. But for non-zero angular momentum test particles, the angular velocity is given by equation (\ref{eqn3}). And we conclude that $-aE$ indicates the space-time dragging term.

It is visible that the dragging is directly proportional to the energy of the particles. Referring to equation (\ref{eqn17}), the space-time forbids the space-time dragging effect on angular momentum-less massive (timelike) particles outside $ar=1$. As one moves outward from the axis, the effect gradually increases and suddenly vanishes at $ar=1$. So $ar=1$ is also the boundary of the space-time dragging effect for angular momentum-less test particles.

\section{Closed Timelike Orbits in Van Stockum space-time}\label{closed_orbit}

The Van Stockum dust solution contains a causality violating region and hence, allows closed timelike curves (CTC). The CTC in the interior solution was found in the circular timelike $\phi$ coordinate with $t,r,z$ being constants. It appears at the $ar>1$ region with $r<R$ (where $R$ is the radius of the dust cylinder). The condition improvises equation (\ref{eqn1}) to be,
\begin{equation}\label{eqn28}
	ds^2=r^2(1-a^2r^2)d\phi^2.
\end{equation}

Now we use Lagrangian of the system and obtain the canonical momentum of $\phi$ coordinate i.e. angular momentum of the test particle revolving in CTC,
\begin{equation}\label{eqn29}
	\dot{\phi}=\frac{P_{\phi}}{r^2(1-a^2r^2)}.
\end{equation}

Interestingly this equation does not involve any separate space-time dragging term. Here the angular velocity of the test particle only involves spatial angular momentum term. So in the vicinity of CTC, angular momentum-less particles do not have non-zero angular velocity as space-time dragging is absent for angular momentum-less particles. Moreover, if $\dot{\phi}$ is zero, $\phi$ becomes constant violating the condition for CTC to occur. So it forbids any angular momentum-less particles to traverse any rotational orbits. Thus they can not be present in closed timelike curves.

Now we initiate our study of closed timelike geodesics by applying the conditions of closed timelike orbits to the geodesics.

One of such conditions is $t,r,z$ to be constant. So it will allow us to consider $\dot{t}=\dot{r}=\dot{z}=0$. From equations (\ref{eqn2}) (\ref{eqn4}) and (\ref{eqn14}),
\begin{equation}\label{eqn30}
	\dot{t}=(1-a^2r^2)E+aP_\phi=0,
\end{equation}
\begin{equation}\label{eqn31}
	\dot{z}=P_z e^{a^2r^2}=0,
\end{equation}
\begin{equation}\label{eqn32}
	\begin{split}
	\dot{r}^2=e^{a^2 r^2} [-1+(1-a^2 r^2) E^2 +2aEP_\phi- \frac{P_\phi^2}{r^2} &\\ -P_z e^{a^2 r^2}]=0&
\end{split}
\end{equation}
as, $e^{a^2 r^2} \neq 0$, so one gets,
\begin{equation}\label{eqn33}
	(1-a^2r^2)E=-aP_\phi,
\end{equation}
\begin{equation}\label{eqn34}
	P_z=0,
\end{equation}
\begin{equation}\label{eqn35}
	-1+(1-a^2 r^2 ) E^2+2aEP_\phi-\frac{P_\phi^2}{r^2}=0.
\end{equation}

Using equations (\ref{eqn33}) and (\ref{eqn34}) to equation (\ref{eqn35}) one gets,
\begin{equation}\label{eqn36}
	E=\pm \frac{ar}{\sqrt{a^2r^2-1}}.
\end{equation}

Therefore the amount of energy necessary for a test particle to traverse within a closed timelike geodesics is obtained. Further, the formula re-established the position of closed timelike geodesics to be $ar>1$.

With the same treatment, the energy of a particle traveling in the closed spacelike geodesics (CSG) can be shown as,
\begin{equation}\label{eqn37}
	E=\pm \frac{ar}{\sqrt{1-a^2r^2}},
\end{equation}
which suggests that CSG is only allowed in $ar<1$ region of the cylinder.

So the $ar=1$ line describes the boundary between CTG and CSG. It is the position where closed null geodesic (CNG) is found to be possible. Also, the line is the boundary of the causality violating region. The $ar=1$ boundary is nothing but the Cauchy horizon.

\subsection{Backward Time Jump in CTG}\label{jump_ctg}
Now using equations (\ref{eqn33}), (\ref{eqn34}) and (\ref{eqn36}) to equation (\ref{eqn3}), one gets,
\begin{equation}\label{eqn38}
	\dot{\phi}=-\frac{E}{ar^2}=\mp \frac{1}{r \sqrt{a^2r^2-1}}.
\end{equation}

Now let $\tau_1$  and $\tau_2$ be the proper time of the test particle at $\phi=0$ and $\phi=2 \pi$ position respectively. During the time reversal, backward time jump occurs at $\phi=2 \pi$ position, and $\Delta \tau=\tau_2-\tau_1$ would be the backward time jump, i.e.
\begin{equation}\label{eqn39}
	2\pi =\mp \frac{1}{r \sqrt{a^2r^2-1}}(\tau_2-\tau_1) =\mp \frac{1}{r \sqrt{a^2r^2-1}}(\Delta \tau).
\end{equation}

Considering non-negative energy of the particle, the formula takes the form,
\begin{equation}\label{eqn40}
	\Delta \tau= -2 \pi r \sqrt{a^2r^2-1}.
\end{equation}

The negative sign indicates the backward time jump.

The formula shows the constant nature of backward time jump (as $a$ and $r$ are both constants), which suggests that after every full rotation the test particle will always travel back to the same proper time position it started the journey. That means for a specific value of ‘$r$’, the particle cannot escape the closed timelike loop where it will always travel for a specific instant of time and will reverse back to the same starting point.

\begin{figure}[h!]
	\centerline{\includegraphics[scale=.6]{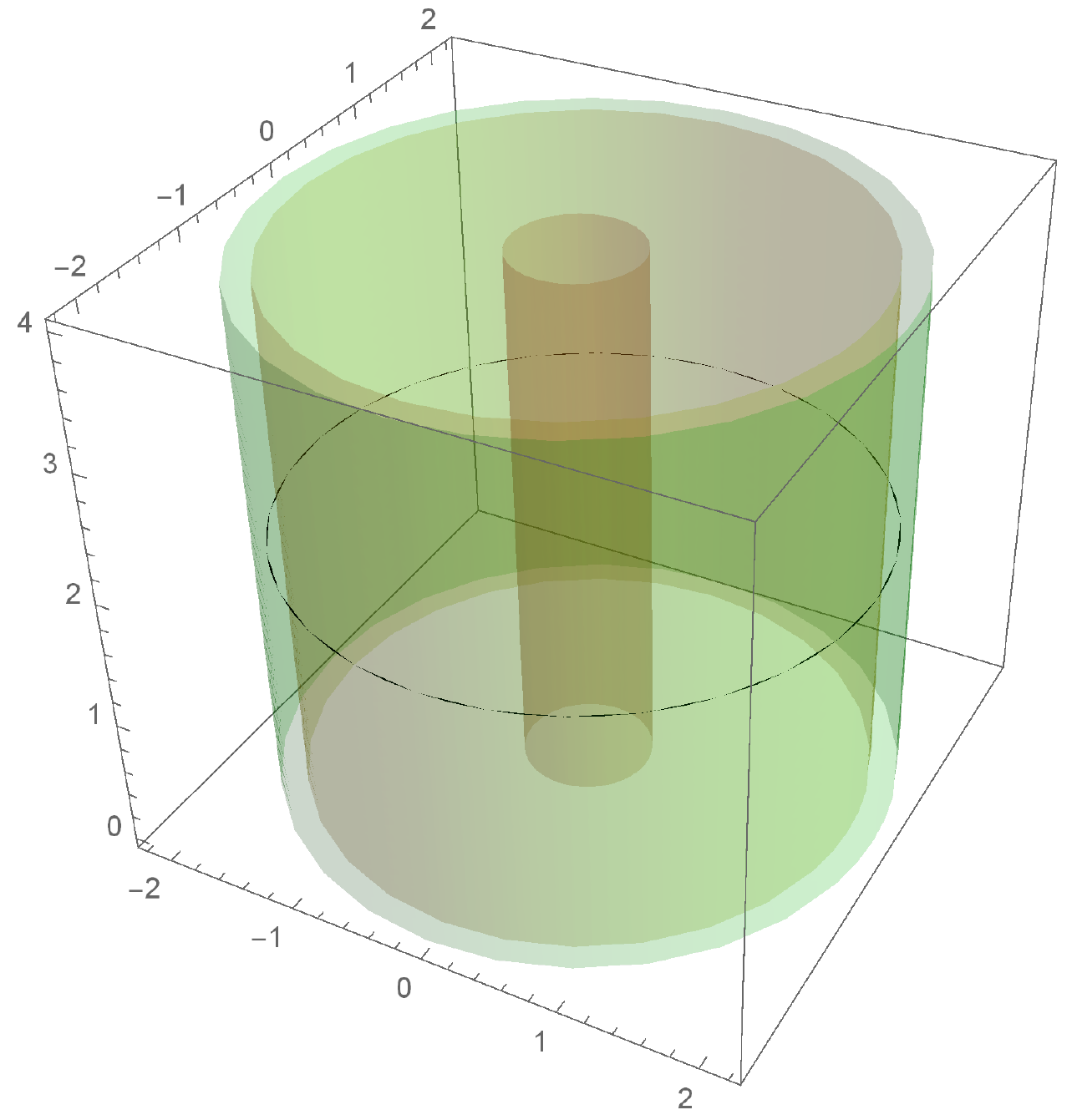}}
	\caption{Parametric plot for $E=2, a=\frac{1}{2}, P_\phi=1$ showing $r_{min-time}$ boundary at $r=0.457$ (inner cylinder), $ar=1$ boundary at $r=2$ (middle cylinder), $r_{max-time}$ boundary at $r=2.189$ (outer cylinder) and a closed timelike geodesic at $r=2.1, z=2$ position (black circular line)}
	\label{plot_ctg}
\end{figure}

\subsection{Dependence on Angular Momentum ($P_\phi$)}\label{angular_dep}
From sections \ref{null_zero-ang} and \ref{time_zero-ang} one may study the movement of an angular momentum-less particle. The particle trajectory is restricted within $ar<1$. So the presence of angular momentum-less particles is forbidden in the $ar>1$ region. But particle with $P_\phi \neq 0$ can be seen in the mentioned region as described in section \ref{time_ang}. So particles with non-zero angular momentum can only traverse in CTG of Van Stockum interior solution. From equation (\ref{eqn33}), one gets,
\begin{equation}\label{eqn41}
	E=- \frac{aP_\phi}{1-a^2 r^2}.
\end{equation}
So without the presence of non-zero angular momentum, the particle has no energy to survive and the whole theory breaks down.

A parametric plot has been employed in figure \ref{plot_ctg} showing the $ar=1$ boundary, $r_{min-time}$, $r_{max-time}$. All the three regions defined in section \ref{region} are visible. A closed timelike geodesic is also shown (black circular line in Region III).

\section{A General Prescription}\label{gen_pres}

Sections \ref{geodesic_motion}, \ref{region}, and \ref{closed_orbit} briefly describe that Van Stockum's interior solution does not allow angular
momentum-less particles to generate CTC. The position of rotating dust where CTC forms restrict the movement of angular momentum-less particles. This leads us to proceed with a general version of the axially symmetric metric to develop a more general concept of the study.

\subsection{Closed Timelike Curves}\label{gen_ctc}

The general stationary axially symmetric rotating line element is defined as,
\begin{equation}\label{eqn42}
	ds^2=-Fdt^2+2Md \phi dt+Ld \phi^2+ H_1dr^2+H_2dz^2,
\end{equation}
where $F$,$M$, $L$, $H_1$ and $H_2$ are metric components which are functions of $r, t, z$ together or individually.
Here the azimuthal curve with $t$, $r$, $z$ constant is a closed curve where $L<0$ defines the orbit as closed timelike curve. Also the $L=0$ describes a closed null curve.

The Lagrangian of the metric is,
\begin{equation}\label{eqn43}
	2\mathcal{L}=-F \dot{t}^2+2M \dot{\phi} \dot{t}+L \dot{\phi}^2+ H_1 \dot{r}^2+H_2 \dot{z}^2.
\end{equation}

We get the first order geodesics by calculating canonical momentum $P_q=\frac{\partial \mathcal{L}}{\partial \dot{q}}$. Hence,
\begin{equation}\label{eqn44}
	P_t=-F \dot{t}+M \dot{\phi}=-E,
\end{equation}
\begin{equation}\label{eqn45}
	P_\phi=M \dot{t}+L \dot{\phi},
\end{equation}
\begin{equation}\label{eqn46}
	P_z=H_2 \dot{z}.
\end{equation}

Here, $E, P_\phi,P_z$ are integration constants termed as total energy of the test particle, angular momentum, and momentum along z direction respectively. The equations now simplify to,
\begin{equation}\label{eqn47}
	\dot{t}=\frac{P_\phi M+EL}{M^2+LF},
\end{equation}
\begin{equation}\label{eqn48}
	\dot{\phi}=\frac{P_\phi F-EM}{M^2+LF},
\end{equation}
and
\begin{equation}\label{eqn49}
	\dot{z}=\frac{P_z}{H_2}.
\end{equation}

Here $\dot{\phi}$ represents angular velocity of the test particle. It is interesting to conclude that angular momentum-less particles still have a non-zero angular velocity given by,
\begin{equation}\label{eqn50}
	\dot{\phi}=\frac{-EM}{M^2+LF}.
\end{equation}

The velocity directly depends on the energy of the test particle. It appears due to the rotation of the space-time and is interpreted as the consequences of space-time dragging. So while studying test particle movements within closed timelike paths we can not simply neglect angular momentum-less particles. Equation (\ref{eqn50}) generalizes equation (\ref{eqn27}).

We initiate the study by calculating the radial geodesic equation. For radial null geodesics, equation (\ref{eqn42}) becomes,
\begin{equation}\label{eqn51}
	H_1 \dot{r}^2=F \dot{t}^2-2M \dot{\phi} \dot{t}-L \dot{\phi}^2-H_2 \dot{z}^2.
\end{equation}

Hence, after using equations (\ref{eqn47}), (\ref{eqn48}) and (\ref{eqn49}) in equation (\ref{eqn51}) we have,
\begin{equation}\label{eqn52}
	\dot{r}^2=\frac{E^2L-P_\phi^2F+2MEP_\phi}{(M^2+LF)H_1}-\frac{P_z^2}{H_1 H_2}.
\end{equation}

Using equations (\ref{eqn47}) and (\ref{eqn52}) radial confinements are found. 

Taking the $z$-axial movement to be free, we have, 
\begin{equation}\label{eqn53}
	\frac{\dot{t}}{\dot{r}}=\frac{dt}{dr}= 
	\frac{(P_\phi M+EL)\sqrt{H_1}}{[(M^2+LF)(E^2L-P_\phi^2F+2MEP_\phi)]^\frac{1}{2}}.
\end{equation}

The denominator must be greater than zero to maintain the equation real. Now as the Lorentzian signature of the metric ($(M^2+LF)>0$) holds for all cases, i.e.
\begin{equation}\label{eqn54}
	(E^2L-P_\phi^2F+2MEP_\phi)>0,
\end{equation}
provides us with the radial geodesic confinement of photons. Therefore, for angular momentum-less photons, $E^2L>0$. As the energy of the particles is always non-zero,
\begin{equation}\label{eqn55}
	L>0,
\end{equation}
is the desired condition to obtain angular momentum-less photon trajectory confinement.

The region where CTC appears (i.e. the causality violating region) is given by $L<0$. Now as the photon confinement boundary creates the timelike particle confinement, the condition always restricts angular momentum-less massive (timelike) particles to travel to the CTC region.

Equation (\ref{eqn55}) perfectly fits with the explanation of equations (\ref{eqn8}) and (\ref{eqn17}) of Van Stockum. It provides the same result as described in section \ref{closed_orbit}.

For CTC, the line element given by equation (\ref{eqn42}) reduces to,
\begin{equation}\label{eqn56}
	ds^2=L d\phi^2.
\end{equation}

Calculating the canonical momentum by using the Lagrangian of the motion, we obtain the angular velocity of the test particles traversing the CTC, as,
\begin{equation}\label{eqn57}
	\dot{\phi}=\frac{P_\phi}{L}.
\end{equation}

Comparing this equation with equation (\ref{eqn48}) confirms that the metric component $M$ must be zero. Hence, it proves that the space-time dragging term given by equation (\ref{eqn50}) is absent in closed timelike curves. We can conclude that CTC may not be the result of space-time dragging.

Moreover, equation (\ref{eqn57}) also verifies that angular momentum-less particles have no angular velocity to orbit in CTC.

\subsection{Closed Timelike Geodesics}\label{gen_ctg}

Now to further analyze CTG, we consider that the space-time metric given by equation (\ref{eqn42}) admits closed timelike geodesics. Hence for timelike geodesics, equation (\ref{eqn42}) becomes,
\begin{equation}\label{eqn58}
	H_1 \dot{r}^2=-1+F \dot{t}^2-2M \dot{\phi} \dot{t}-L \dot{\phi}^2-H_2 \dot{z}^2.
\end{equation}

Using the values of equations (\ref{eqn47}), (\ref{eqn48}) and (\ref{eqn49}) in equation (\ref{eqn58}),
\begin{equation}\label{eqn59}
	\dot{r}^2=\frac{-M^2-LF+E^2L-P_\phi^2F+2MEP_\phi}{(M^2+LF)H_1}-\frac{P_z^2}{H_1 H_2}.
\end{equation}

Now if we impose the conditions of CTC on the timelike geodesic equations (\ref{eqn47}), (\ref{eqn48}), (\ref{eqn49}) and (\ref{eqn59}), the curve defines closed timelike geodesics. According to the conditions, $\dot{t}=\dot{r}=\dot{z}=0$. Therefore we get,
\begin{equation}\label{eqn60}
	P_\phi M+EL=0,
\end{equation}
\begin{equation}\label{eqn61}
	P_z=0,
\end{equation}
and
\begin{equation}\label{eqn62}
	-M^2-LF+E^2L-P_\phi^2F+2MEP_\phi=0.
\end{equation}
Equation (\ref{eqn60}) gives,
\begin{equation}\label{eqn63}
	P_\phi =- \frac{EL}{M}.
\end{equation}

It is clear that if $P_\phi=0$, $E$ will always be zero (since $L$ cannot be zero for timelike curves) implying that the particle has no energy. Again, if we put $P_\phi=E=0$ in equation (\ref{eqn62}) it gives $M^2+LF=0$ which simply contradicts the Lorentzian signature of the space-time. So $P_\phi$ can not be zero, i.e. angular momentum-less particles are forbidden in closed timelike geodesics.

If we put the value of $P_\phi$ from equation (\ref{eqn63}) to equation (\ref{eqn62}) and proceed further maintaining the Lorentzian signature of the metric we get,
\begin{equation}\label{eqn64}
	(-M^2-E^2L)=0.
\end{equation}
Hence,
\begin{equation}\label{eqn65}
	E =\pm \frac{M}{\sqrt{-L}}.
\end{equation}

This is the amount of energy a particle requires to traverse within a CTG.

Again if we put equation (\ref{eqn63}) and (\ref{eqn65}) in equation (\ref{eqn48}) it gives,
\begin{equation}\label{eqn66}
	\dot{\phi}=- \frac{E}{M}=\mp \frac{1}{\sqrt{-L}},
\end{equation}
and integrating,
\begin{equation}\label{eqn67}
	\Delta \tau=\mp 2 \pi \sqrt{-L},
\end{equation}
where $\Delta \tau=(\tau_2-\tau_1)$ is the backward time jump of CTG. $\tau_1$  and $\tau_2$ are the proper time of the massive (timelike) particle at $\phi=0$ and $\phi=2 \pi$ respectively.

The above conditions of (\ref{eqn65}) and (\ref{eqn67}) will be fulfilled by every CTG present in axially symmetric rotating metrics of form stated in equation (\ref{eqn42}). But those may not be fulfilled by the CTCs which are not geodesics in nature. Equations (\ref{eqn63}), (\ref{eqn65}), and (\ref{eqn67}) generalize equations (\ref{eqn41}), (\ref{eqn36}), and (\ref{eqn40}) respectively.

\section{Examples on the General Prescription}\label{gen_example}

We will study a few well-known axially symmetric solutions allowing CTC to apply the general prescription.

\subsection{G\"odel Universe}\label{gen_godel}

The line element in G\"odel's universe is given by \cite{RevModPhys.21.447},
\begin{equation}\label{eqn68}
	\begin{split}
	ds^2=4a^2[dt^2-dr^2+(sinh^4r-sinh^2r)d\phi^2 \\ +2\sqrt{2}sinh^2r d\phi dt-dz^2]
	\end{split}
\end{equation}

Here we consider an alternative set of coordinates ($t', r,\phi$) to suppress irrelevant $z$ coordinate \cite{lobo2010closed}. So one gets,
\begin{equation}\label{eqn69}
	\begin{split}
	ds^2=2\omega^{-2}[-dt'^2+dr^2-(sinh^4r-sinh^2r)d\phi^2 \\ +2\sqrt{2}sinh^2r d\phi dt]
	\end{split}
\end{equation}

The set of metric components in the alternative set of coordinates are
$F(r)=2\omega^{-2}, M(r)=2\omega^{-2} \sqrt{2} sinh^2r, L(r)=-2\omega^{-2}(sinh^4r-sinh^2r), H_1(r)=2\omega^{-2}$.

In his cosmological solution G\"odel described that CTC appears at $r>ln(1+\sqrt{2})$. The CTC present in the solution is always non-geodesic in nature \cite{Chandrasekhar1961} \cite{HawkingEllis1973}. A simple way to check the presence of CTG is described by Gr{\o}n and Johannesen \cite{Gr_n_2008}.

To rule out the possibility of angular momentum-less particles traversing within this CTC, we analyze the confinement of angular momentum-less photon trajectory. It establishes the restriction of massive (timelike) angular momentum-less particle movement outside of the confinement boundary.

The geodesic confinement in G\"odel space is briefly analyzed by Novello et al. \cite{PhysRevD.27.779}.

Now to check the angular momentum-less photon confinement we use equation (\ref{eqn55}),
\begin{equation}\label{eqn70}
	-2\omega^{-2}(sinh^4r-sinh^2r)>0,
\end{equation}
which simplifies to $r<ln(1+\sqrt{2})$. This result infers that the maximum radial reach of angular momentum-less photons are $ln(1+\sqrt{2})$, and it immediately rules out the possibility of angular momentum-less timelike particle movement to the CTC region.

\subsection{Bonnor's Rotating dust}\label{gen_bonnor}

Bonnor studied an axially symmetric, stationary exact solution of Einstein’s equations for rotating dust cloud \cite{Bonnor_1977}. The metric is given by,
\begin{equation}\label{eqn71}
	ds^2=-dt^2+2n d\phi dt+(r^2-n^2)d\phi^2 +e^\mu (dr^2+dz^2),
\end{equation}
where,
\begin{equation}\label{eqn72}
	n=\frac{2hr^2}{R^3},~~~~~ \mu=\frac{h^2r^2(r^2-8z^2)}{2R^8},~~~~~ R^2=(r^2+z^2),
\end{equation}

'$h$' is the rotation parameter. The metric components are $ F(r, z)=1 $, $ M(r, z)=n $, $ L(r, z)=(r^2-n^2) $, and $ H_1(r, z)=H_2(r, z)=e^\mu $. The solution admits closed timelike curves which are non-geodesic \cite{Collas2004} \cite{Gr_n_2008}.

The causality violating region where CTC appears can be found by the condition $L<0$. At $z=0$ position the condition becomes,
\begin{equation}\label{eqn73}
	r^4<4h^2.
\end{equation}

Now the angular momentum-less photon confinement at $z=0$ is found with $L>0$, and it takes the form,
\begin{equation}\label{eqn74}
	r^4>4h^2.
\end{equation}

Hence it proves the absence of angular momentum-less massive (timelike) particle movement in the neighborhood of closed timelike curves.

\section{Conclusion}\label{conclusion}
1. Opher et al. observed that in Van Stockum, "The gravitational collapse of a cylinder with rotation can never develop singularities at the axis". It is observed in section \ref{region} of the current study that any particle trapped inside the dust cylinder can never escape (considering the radius of the dust cylinder $R$ is always greater than $r_{max}$). The dust cylinder may not obey the characteristics of a black hole, but it will always appear as a perfect black cylindrical region to the outside observers.
\newline
2. There are some possibilities of unphysical imaginary timelike geodesic solutions in Van Stockum. But imposing the same conditions we found physically realizable null trajectories, and here the unique phenomenon happens, where only photons are allowed to exist. We called it a pure null region. Maybe this can be used to describe unique astrophysical phenomena.
\newline
3. For photons with $E>\frac{P_\phi}{2r}$ (where $r$ is the radial position of $ar=1$ boundary) in Van Stockum space, $ar=1$ boundary always lies between $r_{min}$ to $r_{max}$, and with $E=\frac{P_\phi}{2r}$, $r_{min}$ coincides with the boundary $ar=1$. For $E<\frac{P_\phi}{2r}$ the boundary $ar=1$ is no longer confined between $r_{min}$ to $r_{max}$. The cylindrical shell with radius $r_{min}$ to $r_{max}$ shift rightwards of the boundary $ar=1$. The last two cases only realised in pure null region.
\newline
4. Confinement of null trajectory confirms the confinement of both the geodesic and non-geodesic massive (timelike) particle movement. So while the angular momentum-less photon has the confinement within $ar=1$ in Van Stockum, it assures that timelike particles never cross the boundary. The result is verified in equation (\ref{eqn55}).

The causality violating region (i.e. $ar>1$) extends from $ar=1$ to $r_{max}$. This is nothing but Region III as described in section \ref{region}. It only possesses non-zero angular momentum particles. So here it is verified once again that angular momentum-less particles are completely forbidden in CTC. It is inferred that angular momentum is an important parameter for identifying particles to travel within a closed timelike path. 
\newline
5. In closed timelike geodesics the absence of angular momentum-less particles can be shown more easily. In Van Stockum it is given in equation (\ref{eqn41}) where the general verification is obtained in equation (\ref{eqn63}).
\newline
6. The general description of section \ref{gen_pres} confirms that axially symmetric rotating space-times admitting CTC also admits space-time dragging. This effect depends on the qualitative behaviour of the metric coefficient \cite{Collas2004-2}. We studied this dragging effect on particle motion around the closed timelike orbit. Space-time dragging is the element that drags angular momentum-less particles to perform rotational motion and thus angular velocity arises. But this dragging is forbidden in closed timelike curves as seen from equations (\ref{eqn48}), (\ref{eqn50}), and (\ref{eqn57}). Particles traverse CTC only due to its angular momentum.
\newline
7. Referring to equations (\ref{eqn29}) and (\ref{eqn57}), it is observed that the angular velocity of particles in CTC does not involve separate spacetime dragging terms. It may be the case that the creation of CTC does not depend on space-time dragging. A similar kind of explanation can be found in the study of counter-rotation of CTC \cite{andreka2008} \cite{duan2022}.
\newline
8. We went through a brief study on closed timelike geodesics and found the amount of energy necessary to traverse within CTG and backward time jump. In the general space-time descriptions, these are given in equations (\ref{eqn65}) and (\ref{eqn67}). The time jump is always constant (as '$L$' is constant in equation (\ref{eqn67})) and with each rotation the particle travels back to its same past state where it started the journey. It is like an endless process that never fascinates a time traveler. The jump gets larger as we develop the loop more outside of the axis. We can infer this from equation (\ref{eqn40}) in Van Stockum.
\newline
9. In section \ref{gen_example}, we went through the example of a general prescription for some well-known space-times. It is established that all the space-times of the form stated in equation (\ref{eqn42}) forbid angular momentum-less particles in closed timelike orbits.

\end{document}